\documentclass{aastex6}

\usepackage{graphicx}

\begin{document}

\title{Gauging the Helium Abundance of the Galactic Bulge RR Lyrae Stars
\footnote{Based on observations collected at the European Organisation for Astronomical Research in the Southern Hemisphere under ESO programmes 179.B-2002 and 298.D-5048.}
}

\author{
Marcella Marconi\altaffilmark{1},
Dante Minniti\altaffilmark{2,3,4},
}

\altaffiltext{1}{INAF-Osservatorio Astronomico di Capodimonte, via Moiariello 16, 80131, Naples, Italy.}
\altaffiltext{2}{Depto. de Cs. F\'isicas, Facultad de Ciencias Exactas, Universidad Andr\'es Bello, Av. Fernandez Concha 700, Las Condes, Santiago, Chile.}
\altaffiltext{3}{Millennium Institute of Astrophysics, Av. Vicuna Mackenna 4860, 782-0436, Santiago, Chile.}
\altaffiltext{4}{Vatican Observatory, V00120 Vatican City State, Italy.}

\begin{abstract}
We report the first estimate of the He abundance of the population of  RR Lyrae stars in the Galactic bulge.
This is done by comparing the recent observational data with the latest models.
We use the large samples of ab type RR Lyrae stars found by OGLE IV in the inner bulge and by the VVV survey in the outer bulge.
We  present the result from the new models computed by Marconi et al. (2017), showing that the minimum Period for fundamental RR Lyrae pulsators depends on the He content.
By comparing these models with the observations in a Period versus effective temperature plane, we find that the bulk of the bulge ab type RR Lyrae  are consistent with primordial He abundance $Y=0.245$, ruling out a significant He-enriched population. 
This work demonstrates that the He content of the bulge RR Lyrae is different from that of the bulk of the bulge population as traced by the red clump giants, that appear to be significantly more He-rich.
\end{abstract}

\section{Introduction} 
\label{sec:intro}

RR Lyrae are very old variable stars, with ages  typically $>11$ Gyr \citep[e.g.][]{Walker1989},  traditionally adopted as standard candles and tracers of Population II properties. Within the Milky Way (MW), RR Lyrae stars are typical members of Galactic globular clusters and the field of the Galactic halo but  have also been found in large numbers towards the bulge of the MW. The Galactic bulge is dominated by metal-rich stellar populations, but early spectroscopic measurements concluded that the RR Lyrae in the bulge are metal-poor \citep[see e.g.][]{Smith1984}. \citet{Butler1976} measured a mean metallicity of $[Fe/H] =-0.65 \pm 0.15$ for RR Lyrae in the inner bulge and \citet{Rodgers1977} also reported a low metallicity for 27 RR Lyrae in the Palomar-Groningen fields in the outer bulge. Moreover, \citet{WalkerTerndrup1991}  measured a low mean spectroscopic metallicity with a small dispersion $[Fe/H] =-1.0 \pm 0.16$ for 59 bulge RR Lyrae in the Baade window. Kinematically, the bulge RR Lyrae are also very hot, with low rotation and large velocity dispersion. For example, \citet{Gratton1987} measured a large radial velocity dispersion $\sigma=133 \pm 25$ km/s for 17 bulge RR Lyrae in the Baade window, and more recently \citet{Kunder2017} measured $\sigma=126$ from 947 bulge RR Lyrae.
In summary, the  RR Lyrae in the Galactic bulge are old, metal-poor, and kinematically hot. However, their He abundance has never been measured. Such a measurement would be particularly interesting because, even if  not representative of the whole bulge, RR Lyrae trace the oldest and most metal poor populations found in this Galactic component \citep{Minniti1996}, so that they can help us to understand the very early formation and chemical evolution  of the Milky Way.
 
In fact, the He abundance is very difficult to measure in stars, and spectroscopic stellar He measurements are scarce.
A thorough discussion of helium abundance indicators from photometry is given by \citet{Sandquist2000}. 
The R method, based on the idea proposed by Iben (1968), is the main way to estimate He abundance of old stellar populations like globular clusters and the Galactic bulge \citep[see e.g.][and references therein]{cassisi03}.
\citet{Renzini1986} argued about the importance of the He abundance of the Galactic bulge, finding $Y=0.30-0.35$ from the R method.
This is higher that the primordial He abundance, as expected from Big Bang nucleosynthesis and as measured in the oldest known stellar populations, represented by the globular clusters.
\citet{Terndrup1988} and \citet{Minniti1995} also used the same method to measure the bulge He content, finding a lower He abundance ($Y=0.30\pm 0.05$ and $0.28\pm 0.02$, respectively), but confirming that the bulge is not only metal-rich, but also He-rich. In addition, \citet{Minniti1995} found that there is no significant gradient in the mean bulge He abundance measured from the red clump (RC) giants, in the Galactocentric distance range $0.3<R_G<1.6$ kpc.
\citet{NatafGould2012} argued that the enhanced He enrichment of the Galactic bulge can reconcile the different bulge ages estimated by different groups. 
This field is not devoid of controversy. In fact, \citet{Lee2015}, proposed that  the double structure observed in the bulge RC stars  and interpreted as the X-shape structure of the Galactic bulge \citep{McWilliamZoccali2010,Saito2011}, can be due instead due to bulge stars with enhanced He abundance. \citet{Gonzalez2015} counter argued that this cannot explain the observations of the RC across the whole bulge, and in particular of the disappearence of the double RC feature away from the minor axis. This recent controversy underscores the importance of the measurement of the bulge He abundance also for studies of the structure of the inner Milky Way. 

In order to make some progress, here we use another method, based on the fact that  current nonlinear convective pulsation models predict that the minimum Period for fundamental mode RR Lyrae pulsators (RRab) depends strongly on the He content.
Therefore, we use the observed Period distribution of Galactic bulge RRab in comparison with new models by Marconi et al. (2017, in prep).
The bulge RRab observational data come from Optical Gravitational Lensing Experiment \citep[OGLE IV,][]{Pietrukowicz2015} and the VISTA Variable in the Via Lactea \citep[VVV,][]{Minniti2010} surveys, that mapped from the inner to the outer bulge, allowing us also to explore the presence of possible gradients.



This Letter is organized as follows.
In \autoref{sec:sec2} we describe the data used in this research.
In \autoref{sec:sec3} we present the RR Lyrae models.
In \autoref{sec:sec4} we discuss the results and implications of this work.
Finally, our conclusions are summarized in \autoref{sec:sec5}. \\

\section{THE GALACTIC BULGE RR LYRAE SAMPLES}
\label{sec:sec2}

\begin{figure}[t]
\begin{center}
\includegraphics[height = 10 cm]{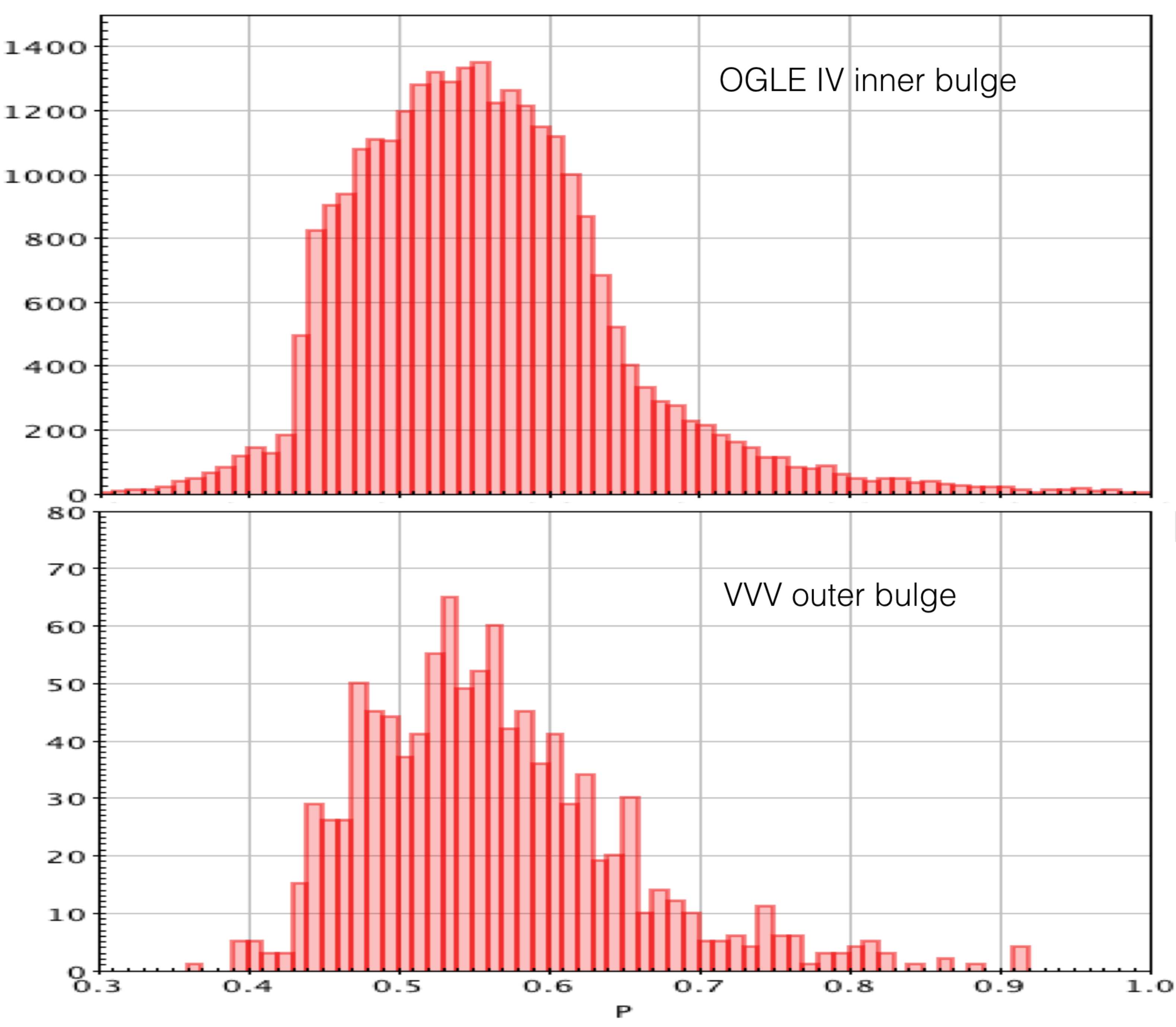}\\
\caption{ 
Period distribution for the Galactic bulge RRab samples of the OGLE IV (top) and the VVV Survey (bottom). }
\label{lc}
\end{center}
\end{figure}

We  use the OGLE IV RRab sample  of 27,258 stars in the inner Galactic bulge ($0.2<R_G<1.4$ kpc) from \citet{Pietrukowicz2015}.  They estimated photometric metallicities following  \citet{Jurksic1995}, \citet{JursicKovacs1996} and \citet{Smolec2005}. A subsample of these OGLE IV variables have VVV near-IR photometry from \citet{Dekany2013}.
We also use the VVV Survey sample of 1019 RRab from \citet{Gran2015}, \citet{Gran2016}. Their RRab search was focused on the outermost tiles of the VVV ($1.1<R_G<2.1$ kpc), where the crowding is not severe and reddening is reduced with respect to the inner regions. 
The VVV observational schedule includes single-epoch photometry in $ZYJHK_s$ bands and variability campaign in $K_s$ band \citep{Minniti2010}. So these near-IR selected RRab are a completely different sample, found in a different way than the optical search made by OGLE, which serves as an excellent comparison to check that the results are not sample dependent.
In Figure 1 we show the period distribution of the Galactic bulge RRab from the OGLE IV (top panel) and the VVV Survey (bottom panel), respectively. The spatial distribution of the two samples is plotted in Figure 2 for a total covered Galactocentric distance range of 0.2 to 2.1 Kpc.

\begin{figure}[t]
\begin{center}
\includegraphics[height = 10 cm]{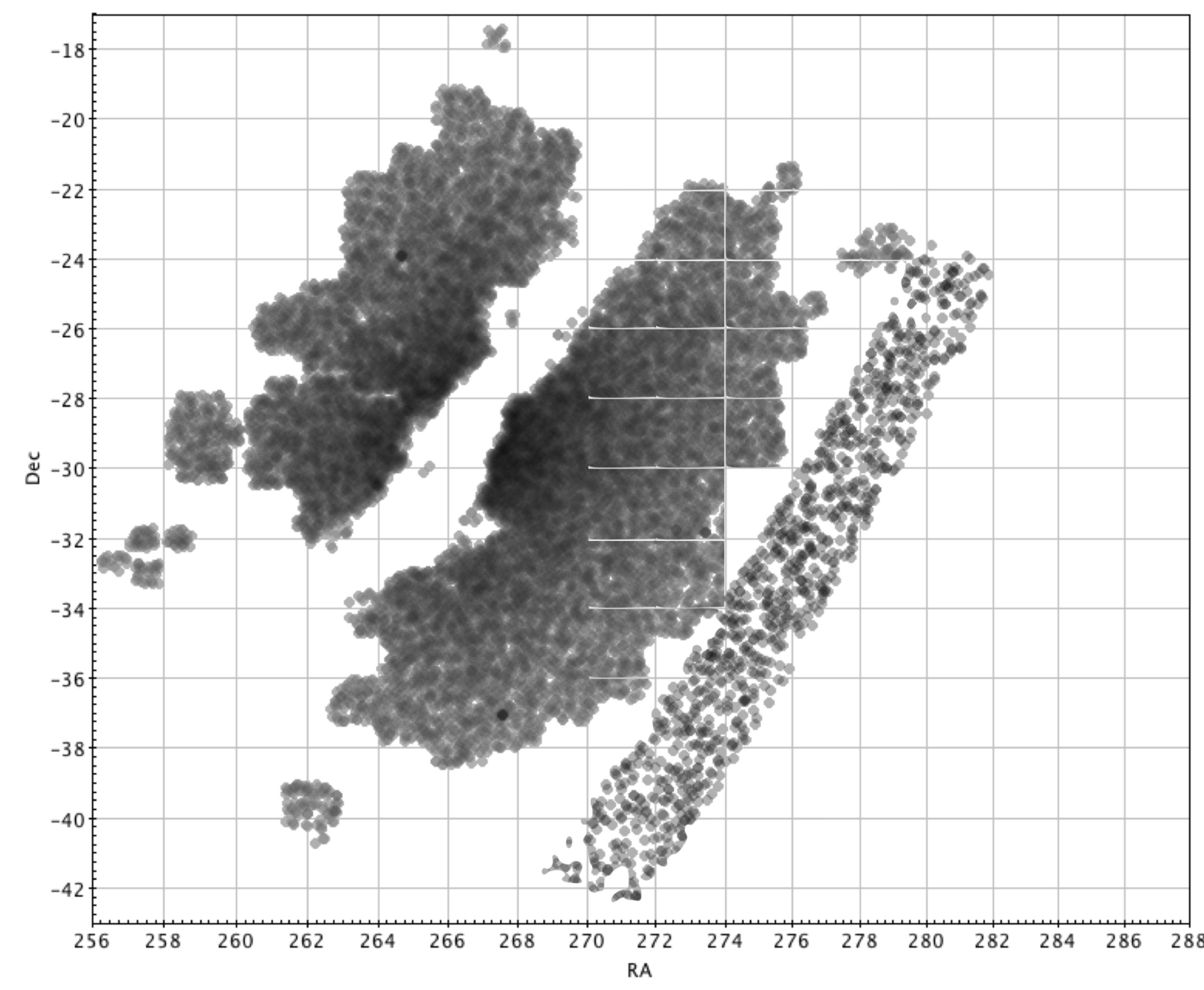}\\
\caption{Spatial distribution of the RRab considered in this work. The inner RRab come from the OGLE IV sample of Pietrukowicz et al. (2015), and the outer bulge RRab come from \citet{Gran2016}. In total the RRab sample covers from $R_G=0.2$ to $2.1$ kpc (adopting a Solar distance of $R_0=8$ kpc).}
\label{map}
\end{center}
\end{figure}

\section{THE MODELS}
\label{sec:sec3}

RR Lyrae are important tracers of Population II stars, and therefore have been subject of extensive modeling, being reasonably well understood not only observationally but also theoretically \citep[e.g.][]{Bono1997,marconi11,marconi15}. 
We have recently computed an updated set of nonlinear convective hydrodynamical models of RR Lyrae stars with the same metal abundances as in \citet{marconi15} but enriched helium content, namely $Y=0.30$ and $Y=0.40$. A grid of these models computed for fundamental and first overtone pulsators for a variety of $Y$ and $Z$ values is shown in Figure 3.
The detailed discussion of the results of these new computations is reported in Marconi et al. (2017, in prep). Here we only anticipate that, in agreement with preliminary models of helium enriched RR Lyrae for the metallicities representative of RR Lyrae in $\omega$ Cen \citep{marconi11}, an increase in the adopted helium abundance corresponds to an increase in the predicted RR Lyrae luminosity levels and in turn to longer predicted periods.

If we consider all the pulsation models computed for $Z$ ranging from 0.0006 to 0.008, roughly representative of Bulge RR Lyrae, namely $[Fe/H]=-1.0\pm0.5$, we can compare the predicted period distribution as a function of the adopted helium abundance and take advantage of the clear dependence of the minimum fundamental period on Y.  This is clearly shown in Figure 3 where first overtone (top panel) and fundamental (bottom panel) models from  Marconi et al. (2017) are plotted  in a period-versus effective temperature diagram, for the labelled helium abundances and metallicity range. In particular, black symbols correspond to standard helium assuming $\Delta{Y}/{\Delta}{Z}=1.4$, with primordial He
abundance of 0.245\citep[see][]{cassisi03,marconi15}, while magenta and blue circles correspond to $Y=0.30$ and $Y=0.40$, respectively.

\begin{figure}
\centering
\includegraphics[height = 10 cm]{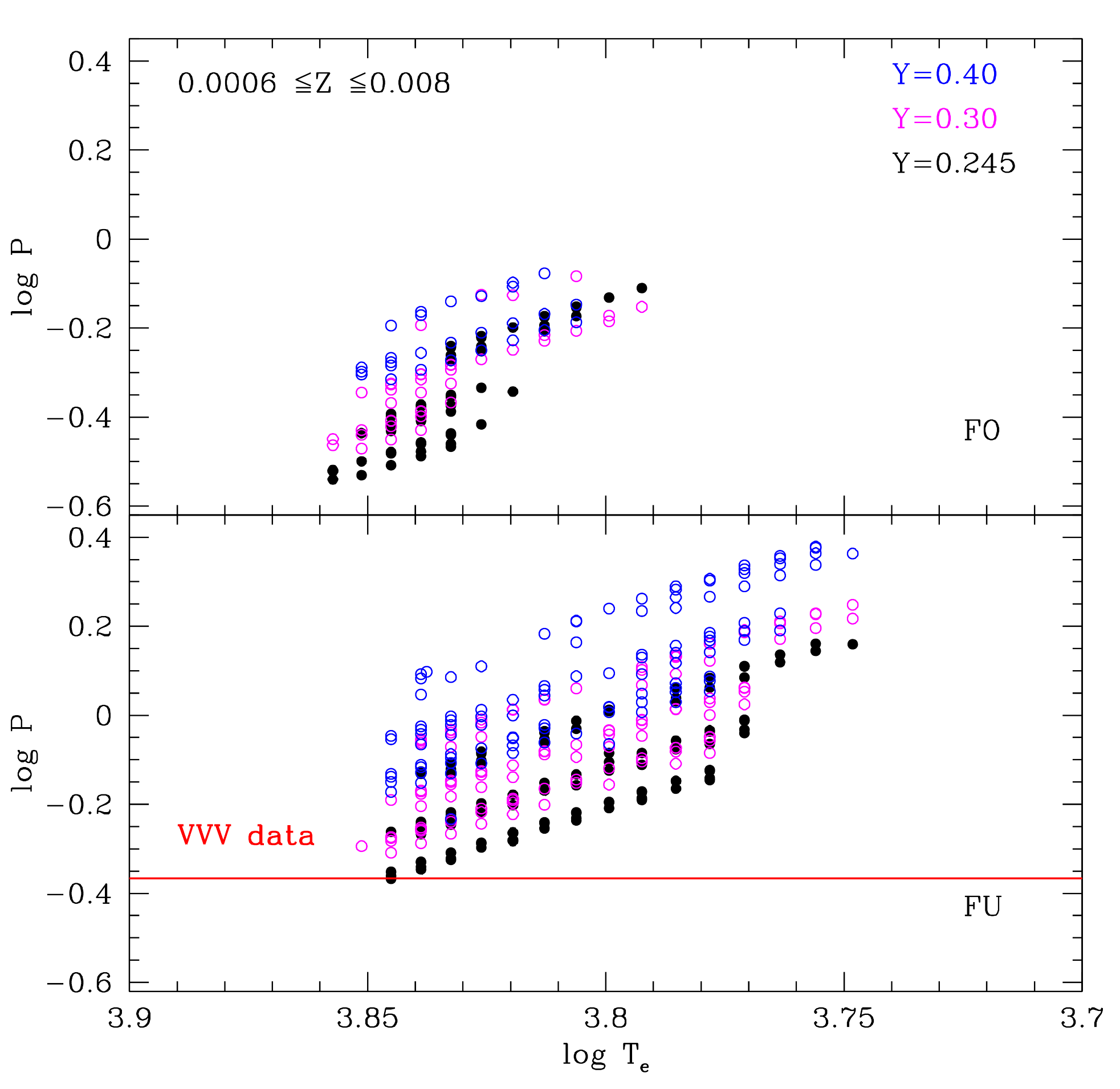}\\
\caption{Periods $vs$ effective temperatures for fundamental RR Lyrae pulsators (bottom) and first overtone pulsators (top), from the models of Marconi et al. (2017) for a range of metallicities and Helium abundances as indicated. We notice that the main effect on period variation is produced by variations in the helium content and the corresponding variation in the luminosity level. Metallicity effects on predicted periods are very small (within few percent). The red line indicates the observed shortest period baseline for comparison.}
\label{fig:timescale}
\end{figure}

\section{RESULTS}
\label{sec:sec4}

As already shown in Figure 1, there is no  significant difference between the VVV and OGLE IV RRab Period distributions.
The minimum RRab period break can be clearly seen at $P=0.430 \pm 0.05$ days,  indicating a similar He content and this period value does not change as a function of Galactocentric distance (over a large range from $R_G=0.2$ kpc out to $R_G=2.1$ kpc).
Moreover, dividing the RRab sample into bins of equal latitude shows no dependence, implying the absence of a significant gradient in the bulge RRab He content (from $b=-10$ to $+5$ deg).
Thus, there appears to be no gradient in the He abundance as a function of Galactic latitude (nor Galactocentric distance).
In the bottom panel of Figure 3 we overplot the measured  minimum RRab period to the fundamental model distribution, suggesting a good agreement between theory and observations for the standard helium abundance.

The top panel of Figure 4 shows the period $vs$ color distribution for the OGLE IV bulge RRab from \citet{Pietrukowicz2015} that have been spectroscopically confirmed by \citet{Kunder2016}.
The near-IR colors for the individual RRab  have been properly dereddened using the maps of  \citet{Gonzalez2011,Gonzales2012}, by taking the mean of a 2 arcmin radius field centered on each target, and assuming the interstellar extinction law from \citet{Cardelli1989}. We note that using the \citet{Nishiyama2009} extinction ratios does not change the results.
The presence of half a dozen redder RRab outliers can be explained by the non-uniform reddening variations.
The  $(V-I)_0$ color can be taken as a proxy for $T_e$, arbitrarily scaled to compare with the models shown in the  middle panel of Figure 4. 
We notice that the predicted trend of increasing period with decreasing effective temperature is not seen so pronounced in the observed period $vs$ color diagram,  as likely due to a significant uncertainty in the dereddened  $(V-I)_0$ colors. Indeed, we verified that when transforming model effective temperatures into $(V-I)_0$ colors, the trend is still predicted.

The thick red line in both panels indicates the shortest period baseline measured for the bulge RRab, $P=0.430 \pm 0.05$ days.  For comparison we show on the right the observed OGLE IV Bulge $RR_{ab}$ period distribution.
This theory $vs$ observations comparison (for mean $[Fe/H]=-1.0 \pm 0.5$ dex) yields a He abundance of $Y=0.245$ for the Galactic bulge RRab, value that is consistent with the primordial He abundance. 
This  value is solid, i.e. it is the same for the two independent datasets (OGLE IV and VVV). 
It is relevant that the existing RR Lyrae spectroscopic metallicities exhibit a small dispersion ($[Fe/H] =-1.0 \pm 0.16$, \citet{WalkerTerndrup1991}. A few ($<10\%$) bulge RRab  appear to be more metal rich (with $[Fe/H]>-0.5$ dex), indicating that the fraction of more He-rich RRab should also be small.
Even though we cannot exclude the presence of a small fraction of stars with $Y=0.30$, we can also conclude that $Y=0.40$ stars are ruled out.
The improvements that can be made for the future are to increase the RRab sample in the inner and outer bulge, and to obtain spectroscopic RRab metallicities.

\begin{figure}
\centering
\includegraphics[height = 15 cm]{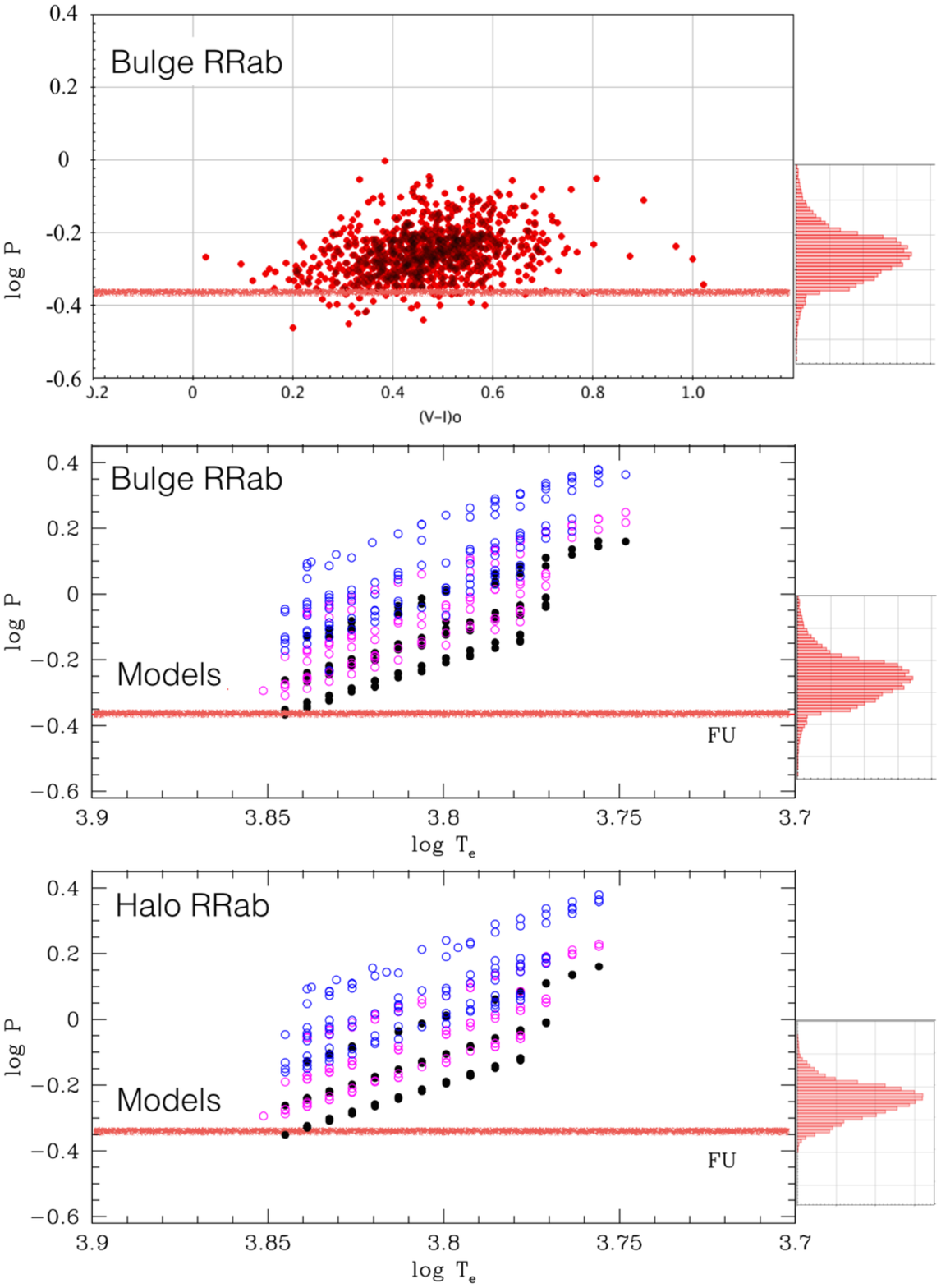}\\
\caption{Comparison of the fundamental RR Lyrae pulsator models from Marconi et al. (2017) from Figure 3 (middle) with the  observations of bulge RRab (top), where the  $(V-I)_0$ color is taken as a proxy for $T_e$ (scaled arbitrarily). 
The thick red line in both panels indicates the shortest period baseline measured for the bulge RRab. 
The period distribution from the bulge RRab of OGLE IV is shown on the right also for comparison. As a check we also show (bottom) the predicted Halo fundamental RR Lyrae  period distribution compared with the histogram of periods from Sesar et al. 2017 on the right.}
\label{fig:timescale}
\end{figure}

In order to check the predictive capabilities of current pulsation models in correctly inferring the helium abundance of a population of RR Lyrae stars, we tested the method against a sample whose helium abundance is generally considered to be known. In particular we selected the PanSTARRS1 catalog of RR Lyrae stars in the Galactic halo by \citet{Sesar2017}, adding the Sloan Digital Sky Survey strip 82 by \citet{Sesar2010}. The helium abundance of these halo RR Lyrae stars is generally considered to be primordial. This result was confirmed by the comparison shown in the lower panel of Figure 4 where the predicted fundamental periods for a metallicity range representative of the Galactic halo population are shown together with the histogram of periods by \citet{Sesar2017} (on the right).
 
\section{Conclusions}
\label{sec:sec5}
We  have presented a  method to estimate the He content of bulge RRab stars, based on the comparison between predicted and observed shortest periods of RRab, that the models indicate are very sensitive to the assumed He content.
We find consistency with the canonical value for old and metal-poor populations of $Y\sim 0.245$  by comparison with the latest models of Marconi et al. (2017).
We cannot exclude the presence of a small fraction of enhanced He abundance RRab, similar to that measured for the bulge RC stars, with mean $Y=0.28-0.35$ \citet{Terndrup1988,Renzini1994,Minniti1995}. However, we can rule out extreme He enhancement ($Y\sim 0.4$) for the bulge RRab stars.
We also find that there is no significant gradient in the bulge RRab He content for a wide range of Galactic latitudes and Galactocentric distances.

The bulge RRab stars appear to be a different population than the bulk of the bulge stars as traced by the RC giants. The more metal-poor and kinematically hot component of the bulge (traced by the RR Lyrae) is probably older and its formation occurred before than the dominating metal-rich component that is traced by the RC giants \citet{Minniti1996,Dekany2013, Kunder2017,Barbuy2017}.
We conclude that the bulk of the bulge RRab are indeed a stellar population with primordial He abundance, as opposed to the bulge RC giants that are He enriched. This is consistent with the bulge RRab being among the oldest stars in the inner MW. 

\acknowledgments
We gratefully acknowledge data from the ESO Public Survey program ID 179.B-2002 taken with the VISTA telescope, and products from the Cambridge Astronomical Survey Unit (CASU). Support is provided by the BASAL Center for Astrophysics and Associated Technologies (CATA) through grant PFB-06, and the Ministry for the Economy, Development and Tourism, Programa Iniciativa Cientifica Milenio grant IC120009, awarded to the Millennium Institute of Astrophysics (MAS). D.M. acknowledges support from FONDECYT Regular grant No. 1170121. We thank G. Bono for a critical reading of the manuscript.\\

\mbox{}

\end{document}